\documentclass[manuscript]{aastex}

\slugcomment{}

\begin{document}


\title{Accelerating universe with time variation of $G$ and $\Lambda$}


\author{F. Darabi\altaffilmark{}}
\affil{Department of Physics Azarbaijan University of Tarbiat Moallem, Tabriz, 53714-161 Iran.\\
Research Institute for Astronomy and Astrophysics of Maragha (RIAAM), Maragha 55134-441, Iran.}
\email{f.darabi@azaruniv.edu}


%

\begin{abstract}
We study a gravitational model in which {\it scale transformations}
play the key role in obtaining dynamical $G$ and $\Lambda$. We take
a scale non-invariant gravitational action with a cosmological
constant and a gravitational coupling constant. Then, by a scale
transformation, through a dilaton field, we obtain a new action
containing cosmological and gravitational coupling terms which are
dynamically dependent on the dilaton field with Higgs type
potential. The vacuum expectation value of this dilaton field,
through spontaneous symmetry breaking on the basis of {\it anthropic
principle}, determines the time variations of $G$ and $\Lambda$. The
relevance of these time variations to the current acceleration of
the universe, coincidence problem,  Mach's cosmological coincidence
and those problems of standard cosmology addressed by inflationary
models, are discussed. The current acceleration of the universe is
shown to be a result of phase transition from radiation toward
matter dominated eras. No real coincidence problem between matter
and vacuum energy densities exists in this model and this apparent
coincidence together with Mach's cosmological coincidence are shown
to be simple consequences of a new kind of scale factor dependence
of the energy momentum density as $\rho \sim a^{-4}$. This model
also provides the possibility for a super fast expansion of the
scale factor at very early universe by introducing exotic type
matter like cosmic strings.
\end{abstract}

\keywords{Time variation of $G$ and $\Lambda$, and acceleration of the
universe}

\section{Introduction}

The question of varying gravitational ``constant'' has been among
the most controversial issues in fundamental physics. It raised by
Dirac who introduced the large number hypothesis \citep[Paper I]{Dira1},\citep[Paper II]{Dira2},\citep[Paper III]{Dira3}, and has recently become a subject of intensive experimental and
theoretical studies \citep{Uzan}. Modern theories, like the
string/M-theory or brane models do not necessarily require such a
variation but they provide a natural and self consistent framework
for such variations by assuming the existence of additional
dimensions. Time variation of couplings in these multidimensional
theories has also recently been studied and their consistency with
the available observational data for distant Type Ia supernovae has
been analyzed in \citep{Lore}. It was shown that in these models a
small variation of gravitational coupling arises that makes distant
supernovae to appear brighter, in contradiction with recent
observations of high $z$ supernovae. However, due to the fact that
the magnitude of the effect is not large enough, one could not
safely discard these multidimensional models. For example, a
positive rate of variation $\frac{\dot{G}(t)}{{G}(t)}$ has been
predicted within a $N=1$ ten-dimensional supergravity and with
non-dynamical dilaton \citep{Wu}.

There are also other models in which the time variation of couplings
is generated by the dynamics of a cosmological scalar (dilaton)
field. For example, Damour {\it et al} have constructed a
generalized Jordan-Brans-Dicke model in which the dilaton field
couples with different strengths to visible and dark matter, and
provided nontrivial bounds on the coupling constants of this field
to matter \citep{Damo}. On the other hand, Bekenstein
\citep{Beken} and Bertolami \citep{Berto}, have introduced
models in which both gravitational coupling and cosmological term
are time dependent. Of particular interest for us is the Bertolami's
results in that the gravitational coupling and cosmological term
respectively behave like $G \sim t$ , $\Lambda \sim t^{-2}$ for one
solution, and $G \sim t^2$ , $\Lambda \sim t^{-2}$ for another one. More
recently, the time dependent gravitational and cosmological terms in some
cosmological models for dark energy and coincidence problems have been studied
and some interesting results on the statefinder parameters have been obtained
\citep[Paper I]{Jamil1},\citep[Paper II]{Jamil2}, \citep{Jamil3}, \citep{Jamil4}, \citep{Jamil5} .

Conformal invariance, on the other hand, has played a key role in
the study of $G$ and $\Lambda$ varying theories. Bekenstein was the
first who introduced this possibility and tried to resolve
$G$-varying problem \citep{Beken}. Conformal invariance implies
that the gravitational theory is invariant under local changes of
units of length and time. These local transformations relate
different unit systems or conformal frames via space time dependent
conformal factors, and these unit systems are dynamically distinct.
This fact leads to variability of the fundamental constants.
Recently, it is shown that one can use this dynamical distinction
between two unit systems usually used in cosmology and particle
physics to construct a cancelation mechanism which reduces a large
cosmological constant to a sufficiently small value, and study the
effects of this model both on the early and late time asymptotic
behavior of the scale factor in the standard cosmological model. The
idea that gravitational coupling may be the result of a spontaneous
conformal symmetry breaking and its relevance to Mach principle is
also studied in \citep[Paper I]{Bisa1},\citep[Paper II]{Bisa2},
\citep[Paper III]{Bisa3}.

Unlike above models based on conformal invariance and its
spontaneously symmetry breaking, the purpose of present paper is to
study a gravitational model in which {\it scale transformations}
play the key role in obtaining dynamical $G$ and $\Lambda$. A scale
transformation is different from a conformal transformation. A
conformal transformation is viewed as ``stretching" all lengths by a
space time dependent conformal factor, namely a ``unit"
transformation. But, a scale transformation is rescaling of metric
by a space time dependent conformal factor, and all lengths are
assumed to remain unchanged. This kind of transformation is not a
``unit" transformation; it is just a dynamical rescaling (
enlargement or contraction ) of a system. We will take a non-scale
invariant gravitational action with a {\it cosmological constant} in
which gravity couples minimally to a dimensionless dilaton field,
and matter couples to a metric which is conformally related, through
the dilaton field, to the gravitational metric. Then by a scale
transformation, through the dilaton field, we obtain a new action in
which gravity couples non-minimally to the dilaton field and matter
couples to the gravitational metric. The field equations reveal a
cosmological term and a gravitational coupling which are dynamically
dependent on the dilaton field ( or conformal factor ) having a
Higgs type potential. The vacuum expectation value of this dilaton
field, through spontaneous symmetry breaking on the basis of {\it
anthropic principle}, determines the correlated time variation of
$G$ and $\Lambda$. The relevance of these time variations to the
current acceleration of the universe, together with coincidence
problem, Mach's cosmological coincidence and the problems of
standard cosmology usually addressed by inflationary models, are
discussed.

\section{Time variation of $G$ and $\Lambda$}

We start with the following action\footnote{We will use the sign
convention $g_{\mu \nu}=diag(+, -, -, -)$.}
\begin{eqnarray}
S&=&\frac{1}{2\bar{\kappa}^2}\int  \sqrt{-g}\:({\cal
R}-2\bar{\Lambda})d^4x\\ \nonumber
&-&\frac{3}{\bar{\kappa}^2}\int  \sqrt{-g}\:g^{\mu \nu}\nabla_{\mu}\sigma
\nabla_{\nu}\sigma d^4x
+ S_m(e^{2\sigma}{g}_{\mu \nu}),\label{1}
\end{eqnarray}
where Einstein-Hilbert action with metric $g_{\mu \nu}$ is minimally
coupled to a dimensionless dilaton field $\sigma$, and the matter is
coupled to gravity with the metric $e^{2\sigma}g_{\mu \nu}$ which is
conformally related to the metric $g_{\mu \nu}$\footnote{The idea of
coupling matter to conformally related metrics has already been
proposed by some authors \citep{Damo}.}. The parameters
$\bar{\kappa}^2$ and $\bar{\Lambda}$ are gravitational coupling and
cosmological constants, respectively. Variation with respect to
$g_{\mu\nu}$ and $\sigma$ yields
\begin{equation}
G_{\mu \nu}+ \bar{\Lambda} g_{\mu \nu}=\bar{\kappa}^2\tilde{T}_{\mu
\nu} + \tau_{\mu \nu},\label{3}
\end{equation}
\begin{equation}
\Box \sigma=-\frac{\bar{\kappa}^2}{6}\tilde{T}, \label{5}
\end{equation}
where
\begin{equation}
\tilde{T}_{\mu \nu}=\frac{2}{\sqrt{-g}}\frac{\delta
S_m(e^{2\sigma}{g}_{\mu \nu})}{\delta g^{\mu \nu}},\label{6}
\end{equation}
\begin{equation}
\tau_{\mu \nu}=6(\nabla_{\mu}\sigma \nabla_{\nu}\sigma
-\frac{1}{2}g_{\mu \nu}\nabla_{\gamma}\sigma
\nabla^{\gamma}\sigma),\label{7}
\end{equation}
and $\tilde{T}$ is the $g_{\mu \nu}$ trace of the energy-momentum
tensor $\tilde{T}_{\mu \nu}$. Now, we introduce the scale
transformations
\begin{equation}
g_{\mu \nu} \rightarrow \Omega^2g_{\mu \nu}, \label{7'}
\end{equation}
\begin{equation}
\sqrt{-g}\rightarrow \Omega^4 \sqrt{-g}, \label{8}
\end{equation}
\begin{equation}
{\cal R}\rightarrow \Omega^{-2}{\cal
R}+6\Omega^{-3}\nabla_{\mu}\nabla_{\nu}\Omega g^{\mu \nu}.\label{9}
\end{equation}
where $\Omega=e^{-\sigma}$. The action (\ref{1}) then becomes
\begin{eqnarray}
 S&=&\frac{1}{2\bar{\kappa}^2}\int \sqrt{-g}\:({\cal R}\Omega^2+6\Omega\Box \Omega
 -2\bar{\Lambda}\Omega^4)d^4x\\ \nonumber &-&\frac{3}{\bar{\kappa}^2}\int  \sqrt{-g}\:g^{\mu \nu}\nabla_{\mu}\Omega
 \nabla_{\nu}\Omega d^4x + S_m(g_{\mu \nu}),\label{10}
\end{eqnarray}
where gravity couples non-minimally to the dilaton field and matter
couples to the gravitational metric $g_{\mu \nu}$. The field
equations are obtained by variation of (\ref{10}) with respect to
the fields $g_{\mu \nu}$ and $\Omega$ as
\begin{equation}
G_{\mu \nu}+ \Omega^2\bar{\Lambda} g_{\mu
\nu}=\Omega^{-2}\bar{\kappa}^2 {T}_{\mu \nu} + \tau_{\mu
\nu}(\Omega),\label{11}
\end{equation}
and
\begin{equation}
\Box \Omega+\frac{1}{12}({\cal R}-4\bar{\Lambda}\Omega^2)\Omega=0,
\label{12}
\end{equation}
where
\begin{eqnarray}
\tau_{\mu \nu}(\Omega)&=&\Omega^{-1}(3\Box \Omega g_{\mu \nu}- 6
\nabla_{\mu}\nabla_{\nu}\Omega)\\ \nonumber
&+&6\Omega^{-2} (-\frac{1}{2}g_{\mu
\nu}\nabla_{\gamma}\Omega \nabla^{\gamma}\Omega+\nabla_{\mu}\Omega
\nabla_{\nu}\Omega).\label{13}
\end{eqnarray}
One may rewrite equation (\ref{11}) as
\begin{equation}
G_{\mu \nu}+ \Lambda g_{\mu \nu}=\kappa^2 {T}_{\mu \nu} + \tau_{\mu
\nu}(\Omega),\label{14}
\end{equation}
where $\Lambda=\Omega^2\bar{\Lambda}$ and
$\kappa^2=\Omega^{-2}\bar{\kappa}^2$. From Eq.(\ref{12}) we infer
the Higgs type potential for $\Omega$
\begin{equation}
V(\Omega)=\frac{1}{24}({\cal
R}-2\bar{\Lambda}\Omega^2)\Omega^2.\label{15}
\end{equation}
For a given negative Ricci scalar\footnote{In the sign convention
$g_{\mu \nu}=diag(+, -, -, -)$ the Ricci scalar for Robertson Walker
metric is negative.}, a positive $\bar{\Lambda}$ leads to vanishing
minimum for the conformal factor, namely $\Omega_{min}=0$. This case
is a failure of conformal transformation with zero cosmological
constant $\Omega^2\bar{\Lambda}$, so is not physically viable. The
non-vanishing minimum of this potential is obtained for a negative
$\bar{\Lambda}$ as
\begin{equation}
\Omega^2_{min}=\frac{\cal R}{4\bar{\Lambda}}>0. \label{15'}
\end{equation}
Putting this as the vacuum expectation value for $\Omega$ in
Eq.(\ref{14}), we obtain
\begin{equation}
G_{\mu \nu}+ \frac{\cal R}{4} g_{\mu \nu}=\frac{4\bar{\Lambda}
\bar{\kappa}^2}{\cal R} {T}_{\mu \nu}.\label{16}
\end{equation}
Equation (\ref{16}) is the $\Omega$ field vacuum condensation of
Eq.(\ref{14}). Considering the Einstein equation in the form
\begin{equation}
G_{\mu \nu} =8 \pi G {T}_{\mu \nu}+\tilde{\Lambda} g_{\mu
\nu},\label{17}
\end{equation}
where $\tilde{\Lambda}=-\Lambda>0 $, we find that the term
$\frac{4\bar{\Lambda} \bar{\kappa}^2}{\cal R}$ accounts for the
dynamical gravitational coupling $G$, and $-\frac{\cal R}{4}$ plays
the role of a positive dynamical cosmological term
$\tilde{\Lambda}$. The negative Ricci scalar asserts to an open
universe with $k=-1$. As far as the Ricci scalar evolves dynamically
with time, the cosmological term and the gravitational coupling will
change in time in a reciprocal way. In other words, $\Lambda
\kappa^2$ is an invariant of the scale transformation (\ref{7}).
Therefore, in the early epoch of time evolution of the universe the
cosmological term and the gravitational coupling may be very large
and very small, in comparison with their current values,
respectively. At very late times, however, where the cosmological
term is very small the gravitational coupling becomes considerably
large, compared with its initial value. This dynamical behavior for
the cosmological term and the gravitational coupling is capable of
being consistent with the current observations. The obtained value
for the cosmological term, namely $\frac{\cal R}{4}$, is consistent
with the upper bound on the current value of cosmological constant:
In an expanding Robertson Walker universe ( with the radius $a$, see
below ) ${\cal R}$ decreases with the radius $a$ as $\sim a^{-2}$,
so the cosmological term in Eq.(\ref{16}), (\ref{17}) fits with its
current observational bound provided $\Omega^2_{min} =\zeta a^{-2}$ where
$\zeta$ is a dimensional constant. This form of decaying $\Lambda$ has already been reported elsewhere \citep{Berto}, \citep{Jafa}, \citep{Chen}, \citep{Silv1}, \citep{Silv2}. The gravitational coupling also
becomes a dynamical quantity with increasing value, in such an
expanding universe. It is easy to arrange for such a small value of
$\bar{\kappa}^2$ so that $\frac{4\bar{\Lambda} \bar{\kappa}^2}{\cal
R}$ equals to the current value of Newtonian gravitational constant
$G$. In fact, for a late time asymptotic power law expansion of the
universe $a(t) \sim t^{\alpha}$ with $\alpha \geq 1$ the time
dependence of $G$ as ${\cal R}^{-1}\sim a^2(t) \sim t^{2\alpha}$
leads to
\begin{equation}
\frac{\dot{G}(t)}{{G}(t)} \sim t^{-1}.\label{18}
\end{equation}
This positive time variation of $G$ has already been reported within
Kaluza-Klein, Einstein-Yang-Mills, Brans-Dicke and
Randall-Sundrum-type models \citep{Lore},\citep{Berto}.

Time variation of $G$ and $\Lambda$ has simple explanation in this
model: The original action (\ref{1}) contains dimensional constants
$\bar{\Lambda}, \bar{\kappa^2}$. By a scale transformation we obtain
the new action (\ref{10}) with dynamical terms ${\Lambda},
{\kappa^2}$. Finally, using the vacuum expectation value of the
$\Omega$ field we obtain the desired time variation for
${\Lambda}(t), {\kappa^2}(t)$. If we were to impose conformal
transformation instead of scale transformation then $\bar{\Lambda},
\bar{\kappa^2}$ would also conformally transform according to their
dimensions as
\begin{equation}
\bar{\Lambda} \rightarrow \Omega^{-2} \bar{\Lambda}, \:\:\:\:\:
\bar{\kappa^2}\rightarrow \Omega^{2} \bar{\kappa^2}.\label{19}
\end{equation}
Eqs.(\ref{11}), (\ref{12}) would then become
\begin{equation}
G_{\mu \nu}+ \bar{\Lambda} g_{\mu \nu}=\bar{\kappa}^2 {T}_{\mu \nu}
+ \tau_{\mu \nu}(\Omega),\label{20}
\end{equation}
and
\begin{equation}
\Box \Omega+\frac{1}{12}({\cal R}-4\bar{\Lambda})\Omega=0,\label{21}
\end{equation}
respectively, which apparently are very different from their
original forms (\ref{14}), (\ref{15}) with completely different
physics. In fact, there is no non-trivial vacuum expectation value
for $\Omega$ except $\Omega_{min}=0$. Therefore, unlike other models
\citep[Paper I]{Bisa1},\citep[Paper II]{Bisa2},
\citep[Paper III]{Bisa3}, in this model only scale transformation can provide
the desired features we are looking for, such as dynamical $\Lambda$
and $\kappa^{2}$.

Vacuum condensation of $\Omega$ field to $\Omega_{min}$ is of
particular importance. As far as $\bar{\Lambda}>0$, there is no
spontaneous symmetry breaking and vacuum condensation. It happens
once $\bar{\Lambda}$ becomes negative. But $\bar{\Lambda}$ is
assumed to be a fundamental constant in the model and can not change
from a positive to a negative value. To resolve this problem one may
resort to the {\it anthropic principle}. According to this
principle, what we perceive as the cosmological constant is in fact
a stochastic variable which varies on a larger structure ( a
multi-verse ), and takes different values in different universes.
Therefore, we live in a universe in which the cosmological constant
is compatible with the development of life. Using this approach in
the present model, one may assume that only those universes with
$\bar{\Lambda}<0$ are capable of condensating the dilaton field to
its vacuum expectation value. Other universes with $\bar{\Lambda}>0$
are then ruled out by anthropic principle. This mechanism now works
like spontaneous symmetry breaking so that discarding
$\bar{\Lambda}>0$ universes in favor of $\bar{\Lambda}<0$ ones,
according to anthropic principle, acts like a selection rule that
causes $\bar{\Lambda}$ to play the role of an {\it effective} order
parameter changing from positive to negative values, and leading to
vacuum condensation of dilaton field in the universes with $\bar{\Lambda}<0$ .

\section{Acceleration of the universe}

We take $g_{\mu \nu}$ and $T_{\mu \nu}$ to be the Robertson-Walker
metric
\begin{equation}
ds^2=c^2dt^2-a^2(t)\left(\frac{dr^2}{(1-kr^2)}+r^2(d\theta^2+\sin^2\theta
d\phi^2 )\right), \label{22}
\end{equation}
and perfect fluid
\begin{equation}
{T}_{\mu \nu}=(\rho+p){u}_{\mu} {u}_{\nu}-p{g}_{\mu \nu}, \label{23}
\end{equation}
respectively, where $k=0 , \pm1$ and $a$ is the scale factor.
Substituting $g_{\mu \nu}$ and $T_{\mu \nu}$ into the Einstein
equation (\ref{17}), provided
$\tilde{\Lambda}=-\bar{\Lambda}\Omega^2_{min}=-\zeta\bar{\Lambda}a^{-2}$,
we obtain the following field equations
\begin{equation}
\frac{\dot{a}^2}{a^2}+\frac{k}{a^2}+\frac{\zeta\bar{\Lambda}}{3a^2}=\frac{1}{3}\bar{\kappa}^2a^2\rho,\label{24}
\end{equation}
\begin{equation}
2\frac{\ddot{a}}{a}+\frac{\dot{a}^2}{a^2}+\frac{k}{a^2}+\frac{\zeta\bar{\Lambda}}{a^2}=-\bar{\kappa}^2a^2p,\label{25}
\end{equation}
where $\dot{a}$ means time derivative with respect to $ct$.
Combining Eqs.(\ref{24}) and (\ref{25}) we obtain the acceleration
equation
\begin{equation}
\frac{\ddot{a}}{a}=-\frac{1}{2}\bar{\kappa}^2a^2(\frac{1}{3}\rho+p)-\frac{\zeta\bar{\Lambda}}{3a^2},\label{26}
\end{equation}
and the conservation equation
\begin{equation}
\dot{\rho}=-3\frac{\dot{a}}{a}\left[(\frac{5}{3}\rho+p)+\frac{2\zeta\bar{\Lambda}}{3\bar{\kappa}^2a^4}
\right].\label{27}
\end{equation}
If we put the power law behavior $\rho=A a^{\alpha}$ and equation of
state $p=\omega \rho$ into Eq.(\ref{27}) the density and pressure of
the perfect fluid are obtained
\begin{equation}
\rho=-\frac{2}{1+3\omega}\bar{\kappa}^{-2}\zeta\bar{\Lambda}a^{-4},\label{28}
\end{equation}
\begin{equation}
p=-\frac{2\omega}{1+3\omega}\bar{\kappa}^{-2}\zeta\bar{\Lambda}a^{-4}.\label{29}
\end{equation}
These results are novel in that the behavior of the density $\rho$
and pressure $p$ in terms of the scale factor $a$ is the same
$a^{-4}$ regardless of the equation of state parameter $\omega$. In
fact, unlike the usual standard cosmology where the equation of
state plays the role of setting the power of the scale factor, it
just sets the multiplicative factors
$-\frac{2}{1+3\omega}\bar{\kappa}^{-2}\zeta\bar{\Lambda}$,
$-\frac{2\omega}{1+3\omega}\bar{\kappa}^{-2}\zeta\bar{\Lambda}$ in this
model, and leaves the power law behavior $\sim a^{-4}$ to be the
same for any equation of state. This is surprising, and tells us
that all types of matter including radiation show the same behavior
with respect to the scale factor, and this brings new opportunities
to solve some cosmological problems, as will be discussed later.

We now put Eqs.(\ref{28}), (\ref{29}) into the acceleration equation
(\ref{26}) which leads to
\begin{equation}
\frac{\ddot{a}}{a}=0,\label{30}
\end{equation}
for each constant value of $\omega$. The energy equation (\ref{24})
with $k=-1$ ( ${\cal R}<0$ ) becomes
\begin{equation}
\dot{a}^2=1-\zeta\frac{\bar{\Lambda}}{3}(1+\frac{2}{1+3\omega}),
\label{31}
\end{equation}
which shows a constant value of $\dot{a}$ for each constant
parameter $\omega$. Integration of this equation, in terms of the
time parameter $ct$ leads to
\begin{equation}
a(t)=a(0)+\sqrt{1-\zeta\frac{\bar{\Lambda}}{3}(1+\frac{2}{1+3\omega})}ct.\label{32}
\end{equation}
Comparing the radiation ($\omega=1/3$) and dust ($\omega=0$)
parameters, we realize that
$$
\rho_{_{Radiation}} < \rho_{_{Dust}}
$$
for a given scale factor. On the other hand,
$$
\dot{a}_{_{Radiation}}< \dot{a}_{_{Dust}.}
$$
This is very interesting because it predicts an acceleration during
the phase transition from radiation to dust eras. The energy
equation also shows interesting features for the early universe: It
is shown by quantum cosmological consideration that the decaying
cosmological term $\Lambda \sim a^{-2}$ may have its origin in the
cosmic strings as a type of exotic matter with the effective
equation of state $p \approx -\frac{1}{3}\rho$ \citep{Jafa}. If we assume such an exotic type of matter would exist in the very early
universe with $\omega\approx -\frac{1}{3}$, then $\dot{a}$ would
become such a huge velocity which could solve the problems of
standard cosmology, such as horizon, flatness and magnetic monopole
problems, in a way similar to the inflationary models.

If $\omega$ would be a continuously varying parameter, the
deceleration parameter would then result in
\begin{equation}
q=-\frac{a \ddot{a}}{\dot{a}^2}=-\frac{\zeta\bar{\Lambda}a
\dot{\omega}}{\dot{a}^3(1+3\omega)^2}.\label{33}
\end{equation}
In principle, $\omega$ has not such a continuous feature; instead it
changes just during phase transitions, and it is reasonable to assume that
the universe decelerates or accelerates because of phase transitions.
We suppose some typical time variation of the parameter $\omega$ during different phase transitions. Then, for example, during phase transition from some exotic type matter era $(\omega\approx -\frac{1}{3})$ toward the
radiation era $(\omega\approx \frac{1}{3})$ we have
$\delta{\omega}>0$ and $q>0$ which accounts for a decelerating
universe. On the other hand, a phase transition from the radiation
$(\omega\approx \frac{1}{3})$ toward dust $(\omega\approx 0)$ eras
leads to $\delta{\omega}<0$ and $q<0$ which shows an accelerating
universe.

\section{Coincidence problem and Mach's cosmological coincidence}

It is usually understood, in the standard cosmology, that the matter
density scales with the expansion of the universe as
$$
\rho_M \sim \frac{1}{a^3},
$$
and the vacuum energy density $\rho_V$ is almost constant. So, there
is only one epoch in the history of the universe when $\rho_M \sim
\rho_V$. It is difficult to understand why we happen to live in this
special epoch. In other words: How finely-tuned is it that we exist
in the era when vacuum and matter are comparable? This is known as
{\it Coincidence problem} \citep{Carr}.

In the present model, the energy density of the vacuum is given by
$\rho(t)=\Lambda(t)/8 \pi G(t)$. Keeping the dimensions of all
quantities, suppose we take $-\bar{\Lambda} \sim 0.1 - 1$. Then,
$\Omega_{min}$ behaves numerically as $\simeq a^{-2}$ which results
in the desired behavior $\Lambda(t) \simeq a^{-2}$, in good
agreement with observational bound on the current value of the
cosmological constant. On the other hand, $G(t) \sim \bar{\kappa}^2
a^2$. Therefore, we obtain the following relation
$$
\rho_V \sim \frac{1}{\bar{\kappa}^2 a^4}.
$$
Unlike the behavior $\rho_M \sim a^{-3}$ in the standard cosmology,
the matter density in the matter dominated era $\omega \approx 0$ ,
according to (\ref{28}), scales with the expansion of the universe
as
$$
\rho_M=-\frac{2\zeta\bar{\Lambda}}{\bar{\kappa}^{2}a^{4}} ,
$$
which has the same scale factor dependence as that of the vacuum
energy density $\rho_V$. Now, demanding $\frac{\rho_M}{\rho_V}
\approx \frac{^0/_030}{^0/_070}$, to account for the current
observation, requires simply
$$
\zeta\bar{\Lambda} \approx -\frac{3}{14}.
$$
Therefore, we come to an important conclusion: There is no real
{coincidence problem}. It is just a result of $\rho_M \sim a^{-4}$
law, in the present nonstandard model. This law, if combined with the mass definition
$$
\rho_M \sim \frac{M}{a^3},
$$
leads to
$$
M \simeq \frac{1}{\bar{\kappa}^2 a}.
$$
But, using $G \simeq \bar{\kappa}^2 a^2$, this relation casts in
the form of the well-known {\it cosmological coincidence} usually
referred to Mach
$$
\frac{GM}{a} \sim 1.
$$
Therefore, the relation ( not coincidence ! ) $\rho_M \sim \rho_V$
casts in the form of Mach's {\it cosmological coincidence}
$\frac{GM}{a} \sim 1$. In a reciprocal way, Mach's {\it cosmological
coincidence} is nothing but a simple result of the $\rho_M \sim
a^{-4}$ law.

\section{Structure formation}

In this section we briefly study the structure formation within the context of present cosmological model. The cosmological principle states that the universe is homogeneous and isotropic over
very large scales. However, on smaller scales the universe is composed of local structures formed by galaxies and cluster of galaxies. The origin of
these local structures are supposed to be the quantum fluctuations in the
very early universe. In fact, the small fluctuations measured in the uniform cosmic microwave background (CMB) are the best evidence for the existence of those early perturbations and are the initial imprint of cosmic structures.

The equations for the amplitude of density perturbations are given by
\begin{equation}\label{34}
\ddot{\delta}+2\frac{\dot{a}}{a}\dot{\delta}=\delta\left(4\pi G\rho-\frac{c_s^2k^2}{a^2} \right),
\end{equation}
for the matter dominant, and 
\begin{equation}\label{35}
\ddot{\delta}+2\frac{\dot{a}}{a}\dot{\delta}=\frac{32\pi}{3}G\rho\delta,
\end{equation}
for the radiation dominant eras, where 
$$
\delta=\frac{\bar{\rho}}{\rho},
$$
is the fractional density perturbation with the plane wave behavior $e^{-ik\cdot
r}$ over the background density $\rho$, and 
$$
c_s^2=\frac{\partial P}{\partial \bar{\rho}},
$$
is the sound speed \citep{Peacok}. In both matter and radiation dominant
eras we have $\rho \propto 1/t^4$. If we take $\delta \propto t^n$, and
set $a(0)\simeq 0$ in (\ref{32}) for simplicity, then we obtain the following equations
\begin{equation}\label{36}
n^2+n+\frac{\bar{\Lambda}}{(1-\zeta\bar{\Lambda})c^2}+c_s^2k^2=0, 
\end{equation}
for the matter dominant, and
\begin{equation}\label{37}
n^2+n+\frac{\bar{\Lambda}}{(1-\frac{2}{3}\zeta\bar{\Lambda})c^2}+c_s^2k^2=0,
\end{equation}
for the radiation dominant eras. Therefore, real valued power $n$ of the growing modes of perturbation, which are given in terms of $\bar{\Lambda}, c, c_s^2, k$, may arise provided we have
\begin{equation}\label{38}
1-\frac{4\bar{\Lambda}}{(1-\zeta\bar{\Lambda})c^2}-4c_s^2k^2\geq0,
\end{equation} 
for the matter dominant, and
\begin{equation}\label{39}
1-\frac{4\bar{\Lambda}}{(1-\frac{2}{3}\zeta\bar{\Lambda})c^2}-4c_s^2k^2\geq0,
\end{equation} 
for the radiation dominant eras.

\section{Nucleosynthesis of light elements}

The abundance of light elements which are resulted in the early universe reactions may be fixed by considering the numbers of neutrons $N_n$ and protons
$N_p$ in an equilibrium with temperature $T$, as the following expression
\begin{equation}\label{40}
\frac{N_n}{N_p}=\exp(-\Delta mc^2/kT).
\end{equation} 
The reason why the neutrons exist today is that the time scale for the weak
interactions required to keep this equilibrium set-up eventually has become
longer than the expansion timescale of the universe at radiation dominant
era. The reactions have 
rapidly ceased and the neutron-proton ratio has freezed-out at some characteristic
value. The quantum field calculations of the effective cross-section for the weak interactions reveals that the interaction timescale changes as $T^{-5}$.
On the other hand, for the radiation dominant era we have
\begin{equation}\label{41}
t=\sqrt{\frac{3}{32\pi G \rho}}.
\end{equation} 
In the standard model of cosmology where $G$ is a constant and $\rho \sim
a^{-4}$, we have the following expansion timescale
$$
t \sim T^{-2},
$$ 
where use has been made of the adiabatic behavior $T\propto a^{-1}$. Therefore, the $T^{-2}$ dependence of the expansion timescale is much slower than the interaction timescale, namely $T^{-5}$, so there is a quite sudden transition between thermal equilibrium and freeze-out. This suggests that the weak interactions switch off and the neutron abundance freezes-out at a temperature of $10^{10.142}K$
leading to an equilibrium neutron-proton ratio of $\frac{N_n}{N_p}\simeq
0.34$. 

We may study this issue within the context of the present non-standard
cosmology. To this end, using $G(t) \sim \bar{\kappa}^2a^2$, 
$\rho=-\bar{\kappa}^{-2}\zeta\bar{\Lambda}a^{-4}$ and $T\propto a^{-1}$, we find that the expansion time scale behaves like $t \sim T^{-1}$. This means that, the expansion timescale in this model is
slower than the one obtained in the standard cosmology, namely $\sim T^{-2}$. Therefore, in comparison with the standard cosmology, a more sudden transition occurs between thermal equilibrium and freeze-out, and the weak interactions are then switched off a little faster than the case in the standard cosmology. This may cause the equilibrium neutron-proton ratio to be a little different
from the one predicted by standard cosmology. However, since the predicted value of the standard cosmology is still far from the experimental value $\frac{N_n}{N_p}\simeq1/6$, one may hope to improve these values by considering some yet unknown phenomenon at very early universe.

\section*{Conclusion}

In this paper, we have studied a model of scale transformation
imposed on a gravitational action which is not scale invariant due
to the presence of dimensional gravitational an cosmological
constants. By choosing a dilaton field for scale transformation, we
obtained a {\it new} action whose field equations revealed dynamical
cosmological term and gravitational coupling. Unlike other
approaches \citep[Paper I]{Bisa1},\citep[Paper II]{Bisa2},
\citep[Paper III]{Bisa3} in which the dynamics of $\Lambda$ ( or perhaps $G$ ) is obtained by resorting to a conformal invariant
action where dynamical distinction between two unit systems in
cosmology and particle physics play a key role, in the present
approach the dynamics of $\Lambda$ and $G$ is not due to any scale
invariance. Instead, in this scale non-invariant theory the dynamics
of $\Lambda$ and $G$ is obtained by introducing a dynamical scale
through a dilaton field. Fixing this scale, through the vacuum
expectation value of dilaton field, leads to time variations of
$\Lambda$ and $G$. These time variations lead the vacuum energy
density to be time dependent as $\rho_v=\Lambda(t)/{8\pi G(t)}$.
Therefore, at early universe where $\Lambda$ is so large and $G$ is
too small compared with their current values, the vacuum energy is
huge. At the present status of the universe, however, the vacuum
energy is vanishing, like $a^{-4}$, due to time variations of both
$\Lambda$ and $G$. This solves the cosmological constant problem in
the present model.

We have also studied the relevance of this model to the other
problems of standard cosmology such as current acceleration of the
universe, coincidence problem, structure formation, nucleosynthesis and those usually addressed by inflationary models. We have shown that there is no principal competition between $\rho_v$ , $\rho_M$ to account for the current
acceleration of the universe. Instead, the important role is played
by the variation of $\omega$, the parameter in the equation of
state, from radiation $(\omega\approx \frac{1}{3})$ toward dust
$(\omega\approx 0)$ eras, which leads to an accelerating universe.
Removing the competition between $\rho_v$ , $\rho_M$ and realizing
the variation of $\omega_R \approx \frac{1}{3}$ towards $\omega_M
\approx 0$ as the reason for the current acceleration of the
universe, the coincidence problem fades away. In fact, there is no 
direct relation between the acceleration of the universe and
coincidence problem in the present model. The acceleration of the
universe owes to the variation of $\omega$ from radiation toward
dust eras and has nothing to do with such coincidence as $\rho_M
\sim \rho_V$. Moreover, there is no coincidence problem in this
model, at all. This is because, all the matter, radiation and vacuum
densities goes like $a^{-4}$ and there is no important competition
between them, in this regard.

However, the model is very sensitive to the equation of state with
$(\omega\approx -\frac{1}{3})$ which might have been occurred by an
exotic type matter, namely cosmic strings, at very early universe.
This value of $\omega$ may cause such a huge velocity for the scale
factor that can resolve the well-known problems ( horizon, flatness,
etc. ) usually addressed by inflationary models. It is worth
noticing that the vacuum energy density $\rho_v$ seems not to play
direct role both in the super fast expansion of the early universe (
due to an exotic matter ) and current acceleration of the universe (
due to phase transition from radiation toward dust eras ). However,
according to (\ref{31}) and (\ref{32}), if $\rho_v$ would be zero,
namely $\bar{\Lambda}=0$, neither super fast expansion of the early
universe nor acceleration of the present universe would then occur.

We have shown that the apparent coincidences $\rho_M \sim \rho_V$
and $ \frac{GM}{a} \sim 1 $ are not really coincidences. They are
just simple results of $\rho \sim a^{-4}$ behavior. This behavior
leads to
$$
M \sim \frac{1}{\bar{\kappa}^2 a},
$$
for which we obtain
$$
\frac{\dot{M}}{M}=-\frac{\dot{a}}{a}=-H,
$$
where $H$ is the Hubble constant. It is important to note that, as in the spirit of the steady state theory, the mass $M$ varies with time. However, the time variation of $M$ is consistent with the modified conservation equation (\ref{27}). Of course, this is the price that we paid for the resolution of coincidence problem in the present non-standard cosmology. One may also
interpret the extra term ${2\bar{\Lambda}}/{3\bar{\kappa}^2a^4}$ in the
conservation equation (\ref{27}) as playing the role of dark energy. We point out that the solution (\ref{32}) for the scale factor becomes free of singularity provided we take $a(0)$, for example, to be a nonzero scale of Planck length.

Finally, considering the cosmological consequences of the acceleration during the phase transition from radiation to dust eras, we have studied briefly
the structure formation. It is shown
that the growing modes of perturbations, as the seeds of structures, are possible both in radiation and matter dominant eras. So, it seems the cosmic acceleration during the phase transition from radiation to dust eras does
not drastically affect the structure formation and this cosmological phenomena
may exist in the matter dominant era. This is mainly because the acceleration eventually stops when $\omega$ becomes almost constant in the
matter dominant era and the phase transition ends. We have also studied
the nucleosynthesis of light elements and shown that the equilibrium neutron-proton ratio is a little different from the predicted one by the standard cosmology.

\section*{Acknowledgment}

The author would like to thank the referee for the useful comments. This work has been supported financially by Research Institute
for Astronomy and Astrophysics of Maragha (RIAAM) under research project
NO.1/2363.

\nocite{*}
\bibliographystyle{spr-mp-nameyear-cnd}
\bibliography{biblio-u1}

\end{document}